\documentclass[preprint,aps,prd,showpacs,superscriptaddress,nofootinbib,tightenlines]{revtex4}

\usepackage{hyperref}
\usepackage{amsfonts}
\usepackage{multirow}
\usepackage{mathrsfs}
\usepackage{graphicx}
\usepackage{amsmath}
\usepackage{amssymb}
\usepackage{bm}
\usepackage{bbm}
\usepackage{color}
\usepackage{array}
\usepackage{diagbox}
\usepackage{physics}
\usepackage{float}
\usepackage{makecell}
\allowdisplaybreaks[4]

\def\OMIT#1{}

\def\hlinew#1{%
  \noalign{\ifnum0=`}\fi\hrule \@height #1 \futurelet
   \reserved@a\@xhline}
\usepackage{array}
\newcommand{\PreserveBackslash}[1]{\let\temp=\\#1\let\\=\temp}
\newcolumntype{C}[1]{>{\PreserveBackslash\centering}p{#1}}
\newcolumntype{R}[1]{>{\PreserveBackslash\raggedleft}p{#1}}
\newcolumntype{L}[1]{>{\PreserveBackslash\raggedright}p{#1}}

\newcommand{\nn}{\nonumber}

\newcommand{\beq}{\begin{equation}}
\newcommand{\eeq}{\end{equation}}
\newcommand{\bqa}{\begin{eqnarray}}
\newcommand{\eqa}{\end{eqnarray}}

\newcommand\fverb{\setbox\fverbbox=\hbox\bgroup\verb}
\newcommand\fverbdo{\egroup\medskip\noindent%
			\fbox{\unhbox\fverbbox}\ }
\newcommand\fverbit{\egroup\item[\fbox{\unhbox\fverbbox}]}
\newbox\fverbbox

\makeatletter

\newcommand{\Rmnum}[1]{\expandafter\@slowromancap\romannumeral #1@}
\makeatother

\begin{document}
%\preprint{}
%%%%%%%%%%%%%%%%%%%%%%%%%%%%%%%%%%%%%%%%%%%%%%%%%%%%%%%%%%%%%%%%%%%%%%%%%%%%%%
\title{\mbox{}\\[10pt]
Light quark fragmentation into S-wave fully charmed tetraquark}

\author{Xiao-Wei Bai~\footnote{ xiaoweibai22@163.com}}
\affiliation{School of Physical Science and Technology, Southwest University, Chongqing 400700, China\vspace{0.2cm}}

\author{Yingsheng Huang~\footnote{yingsheng.huang@outlook.com}}
\affiliation{Department of Physics and Astronomy, University of Utah, Salt Lake City, UT 84112, USA}
\affiliation{High Energy Physics Division, Argonne National Laboratory, Argonne, IL 60439, USA}
\affiliation{Department of Physics \& Astronomy,
	Northwestern University, Evanston, IL 60208, USA}

\author{Wen-Long Sang~\footnote{wlsang@swu.edu.cn}}
\affiliation{School of Physical Science and Technology, Southwest University, Chongqing 400700, China\vspace{0.2cm}}
 
%%%%%%%%%%%%%%%%%%%%%%%%%%%%%%%%%%%%%%%%%%%%%%%%%%%%%%%%%%%%%%%%%%%%%%%%%%%%%%
\date{\today}
%%%%%%%%%%%%%%%%%%%%%%%%%%%%%%%%%%%%%%%%%%%%%%%%%%%%%%%%%%%%%%%%%%%%%%%%%%%%%%
\begin{abstract}
		
% In this work, we compute the fragmentation function of light quark into S-wave fully-charmed tetraquarks ($T_{4c}$) in the framework of nonrelativistic QCD (NRQCD),
% at the lowest order in $\alpha_{s}$ and $v$. The results for the fragmentation functions of light quark to fragment into the $T_{4c}$ are presented, 
% we predict the $T_{4c}$ production rate at high transverse momentum $p_{T}$ regime in \texttt{LHC} and \texttt{EIC}. In addition to light quark fragmentation, $T_{4c}$ can also be produced through charm quark and gluon fragmentation. Thus, we have conducted a comparative analysis of these three fragmentation channels. 
% Our studies reveal that the cross section for $T_{4c}$ production from light quark fragmentation is significantly smaller than that from gluon fragmentation but remains more substantial than that from charm quark fragmentation.
% The integrated cross sections from light quark fragmentation can reach several hundreds femtobarns, indicating a substantial potential for $T_{4c}$ event production at the \texttt{LHC}. 
% Moreover, we estimate the integrated cross sections for $T_{4c}$ production at the \texttt{EIC}. It is suggest that the cross sections are relatively small and the prospect of detecting these $T_{4c}$ at the \texttt{EIC} is somewhat challenging.

We compute the fragmentation function of a light quark into S-wave fully-charmed tetraquarks ($T_{4c}$) within the nonrelativistic QCD (NRQCD) framework, at leading order in $\alpha_{s}$ and $v$. We present results for light quark fragmentation into $T_{4c}$ and predict its contribution to $T_{4c}$ production at high transverse momentum ($p_{T}$) at the LHC and EIC. We also compare light quark fragmentation with charm quark and gluon fragmentation channels. Our analysis shows that the production cross section for $T_{4c}$ from light quark fragmentation is smaller than that from gluon fragmentation but larger than that from charm quark fragmentation. 
% The integrated cross sections from light quark fragmentation reach several hundreds of femtobarns, suggesting significant $T_{4c}$ production potential at the LHC. In contrast, the integrated cross sections at the EIC are smaller, making detection of $T_{4c}$ there more challenging.

\end{abstract}
\maketitle

%---------------------------
\section{introduction}
%---------------------------
The discovery of a new resonance, $X(6900)$, around 6.9 $\rm GeV$ in the di-J/$\psi$ invariant mass spectrum was reported by the LHCb collaboration~\cite{LHCb:2020bwg}, and later confirmed by both the \texttt{ATLAS} and \texttt{CMS} collaborations~\cite{ATLAS:2023bft,CMS:2023owd}. 
Since then, the identification of $X(6900)$ has sparked considerable theoretical interest. 
While alternative interpretations, such as charmonia molecules or hybrids, have been explored, $X(6900)$ is widely regarded as a strong candidate for the fully-charmed tetraquark. 
The fully-charm tetraquark (hereafter denoted by $T_{4c}$), with the charm-quark mass $m_{c}$ well above the perturbative threshold,
can be conceptualized as a composite system made of four nonrelativistic charm and anti-charm quarks.

The study of fully-heavy tetraquarks predates the discovery of $X(6900)$ by several decades~\cite{Iwasaki:1976cn,Chao:1980dv,Ader:1981db}. Since its discovery, significant efforts have focused on the mass spectra and decay properties of $T_{4c}$ using various phenomenological models, 
including quark potential models~\cite{Wu:2016vtq,Becchi:2020uvq,Debastiani:2017msn,Bedolla:2019zwg,Lu:2020cns,liu:2020eha,Zhao:2020nwy,Zhao:2020jvl,Giron:2020wpx,Ke:2021iyh,Jin:2020jfc,Wang:2022yes,Wu:2024euj,Bai:2016int,Gordillo:2020sgc,Wang:2019rdo,Wang:2021kfv,Liu:2021rtn,Yu:2022lak,Mutuk:2021hmi,Tiwari:2021tmz,Faustov:2022mvs},
QCD sum rules~\cite{Chen:2020xwe,Wang:2020ols,Yang:2020wkh,Wan:2020fsk,Zhang:2020xtb,Tang:2024zvf}, Lattice~\cite{Hughes:2017xie} and effective field theories~\cite{Sang:2023ncm,Zhang:2023ffe} (see~\cite{Chen:2022asf} and references therein for a comprehensive overview). 
Despite extensive studies on the mass spectra and decay properties of $T_{4c}$, the understanding of its dynamical production mechanism remains challenging. Primary approaches to modeling its production include the color evaporation model~\cite{Maciula:2020wri,Carvalho:2015nqf} and hadron-quark duality relations~\cite{Karliner:2016zzc,Berezhnoy:2011xy}. Recently, several groups have also explored $T_{4c}$ production within the framework of nonrelativistic QCD (NRQCD) factorization~\cite{Feng:2020riv,Feng:2020qee,Zhang:2020hoh,Huang:2021vtb,Zhu:2020xni,Feng:2023agq,Feng:2023ghc,Bai:2024ezn}. 
 
In high-energy collisions, the large-$p_{T}$ production of identified hadrons is primarily governed by the fragmentation mechanism~\cite{Collins:1989gx}. 
Using the NRQCD factorization theorem, the fragmentation functions for gluons and charm quarks to $T_{4c}$ were computed in~\cite{Feng:2020riv,Bai:2024ezn}, where both the $p_T$ distribution and the integrated cross sections for $T_{4c}$ production via gluon and charm quark fragmentation at the LHC were predicted. These studies suggest that a substantial number of $T_{4c}$ events could be generated at the LHC. Furthermore, the production rates for $T_{4c}$ at electron-ion colliders, such as HERA, EIC, and EicC, were also presented in~\cite{Bai:2024ezn}.
% 
% Recently, Celiberto et al.~\cite{Celiberto:2024mab} investigated $T_{4c}$ production in high-energy proton collisions using single-parton collinear fragmentation within a variable-flavor number scheme (VFNS). 
% They derived a novel set of collinear fragmentation functions, which encode initial-scale inputs corresponding to both gluon and charm fragmentation channels, respectively defined within the context of quark-potential NRQCD and spin-physics inspired models. Light and the bottom-quark are simply generated by evolution in their evolution setup, with no corresponding initial-scale inputs. 
Recently, Celiberto et al.~\cite{Celiberto:2024mab} examined the evolution effects of the gluon and charm quark fragmentation channels in the production of $T_{4c}$ in high-energy proton-proton collisions. In contrast to gluon and charm quark fragmentation, the contributions from light and bottom quarks are generated purely through evolution, as they lack corresponding initial-scale inputs.
% NOTE: consider condensing the above statements about Celiberto's work. mentioning only that they discuss the running of fragmentation functions and light quark fragmentation function could benefit them. 
An important aspect of $T_{4c}$ production mechanisms is the potential contribution from light quark fragmentation. While existing studies have focused on gluon and charm quark fragmenting channels, the light quark channel could provide crucial insights into the overall production mechanism, and potentially compete with the charm quark channel contribution.
% say something about how light quark fragmentation might be important
In this work, we aim to investigate the fragmentation function for light quarks to an $S$-wave $T_{4c}$ by applying the NRQCD factorization framework. Subsequently, we use this fragmentation function to predict $T_{4c}$ production cross sections at the LHC and EIC, comparing the results with those from gluon and charm quark fragmentation channels.

The structure of this paper is organized as follows. Section \ref{qcd-factorization-formula} presents QCD factorization formulas for the fragmentation production of $T_{4c}$ at both proton-proton and electron-ion colliders. In Section \ref{theoretical-framework}, we outline the NRQCD factorization formalism for the fragmentation function. The determination of various short-distance coefficients (SDCs) is discussed in Section \ref{determination-sdcs}. Section \ref{phen-discu} is dedicated to the phenomenological analysis and discussion of $T_{4c}$ production at the LHC and EIC. Finally, we conclude with a summary in Section \ref{summary}.
 
\section{QCD Factorization theorem for high $p_{T}$ production of $T_{4c}$ \label{qcd-factorization-formula}}
The inclusive production of a hadron $H$ at large $p_T$ in high-energy collisions is dominated by the fragmentation mechanism. According to the QCD factorization theorem~\cite{Collins:1989gx}, the inclusive production rate of the $T_{4c}$ at large $p_T$ in proton-proton collisions can be written as follows:
%-----------------------------------
\begin{align}
	\label{t4c-cross section}
	\mathrm{d} \sigma\left(p p \rightarrow T_{4c}\left(p_{\mathrm{T}}\right)+X\right) &= \sum_{a,b,i} \int_0^1 \mathrm{d} x_a  \int_0^1 \mathrm{d} x_b \int_{0}^{1} \mathrm{d} z \, f_{a/p}(x_a,\mu) f_{b/p}(x_b,\mu) \nonumber \\
	& \times \, \hat{\sigma}(a b \rightarrow i(p_T/z) + X, \mu)  D_{i \rightarrow T_{4c}}\left(z,\mu\right) + {\cal O}(1/p_T).
\end{align}
%-----------------------------------
Here, $f_{a,b/p}$ represents the parton distribution functions (PDFs) of a proton, $d {\hat \sigma}$ denotes the partonic cross section, $\mu$ is the factorization scale, $z\in [0,1]$ is the momentum fraction carried by $T_{4c}$ relative to the parent parton $i$, and $D_{i \rightarrow T_{4c}}$ indicates the fragmentation function for parton $i$ into $T_{4c}$.
While the gluon and charm quark fragmentation into $T_{4c}$ have been previously computed in~\cite{Feng:2020riv,Bai:2024ezn}, this work aims to complete the theoretical framework by investigating the light quark fragmentation into $T_{4c}$. At the lowest order in $\alpha_s$, the partonic cross sections of relevant partonic channels are given by
%-----------------------------------
\begin{subequations}
	%-----------------------------------
	\label{partonic:cross:section}
	%----------------------------------
	\begin{align}
	%-----------------------------------
	&\dv{\hat{\sigma}_{gg\rightarrow q_i \bar q_i}}{\hat{t}}=
	\frac{1}{6} \frac{\pi \alpha_{s}^{2} }{\hat s^{2}}\left ( \frac{\hat{t}}{\hat{u}}+\frac{\hat u}{\hat t}-\frac{9}{4}\frac{\hat t^{2}+\hat u^{2}}{\hat s^{2}} \right ),
	%-----------------------------------
	\\
	%-----------------------------------
	&\dv{\hat{\sigma}_{q_i \bar q_i\rightarrow q_i \bar q_i}}{\hat{t}}=
	\frac{\pi \alpha_{s}^{2} }{\hat{s}^{2}}\left [ \frac{4}{9}
	\left ( \frac{\hat{s}^{2}+\hat{u}^{2}}{\hat{t}^{2}}+\frac{\hat{t}^{2}+\hat{u}^{2}}{\hat{s}^{2}}
	\right ) -\frac{8}{27}\frac{\hat{u}^{2}}{\hat{s}\hat{t}} \right ], 
	\\
	%-----------------------------------
	&\dv{\hat{\sigma}_{q_i \bar q_i\rightarrow q_j \bar q_j}}{\hat{t}}=
	\frac{4}{9} \frac{\pi \alpha_{s}^{2} }{\hat s^{2}}\left ( \frac{\hat t^{2}+\hat u^{2}}{\hat s^{2}} \right ), 
	%-----------------------------------
	\\
	%-----------------------------------
	&\dv{\hat{\sigma}_{q_i q_i\rightarrow q_i q_i}}{\hat{t}}=\dv{\hat{\sigma}_{\bar{q}_i \bar{q}_i\rightarrow \bar{q}_i \bar{q}_i}}{\hat{t}}=
	\frac{\pi \alpha_{s}^{2} }{\hat{s}^{2}}\left [ \frac{4}{9}
	\left ( \frac{\hat{s}^{2}+\hat{u}^{2}}{\hat{t}^{2}}+\frac{\hat{s}^{2}+\hat{t}^{2}}{\hat{u}^{2}}
	\right ) -\frac{8}{27}\frac{\hat{s}^{2}}{\hat{u}\hat{t}} \right ], 
	\\
	%-----------------------------------
	&\dv{\hat{\sigma}_{q_i q_j\rightarrow q_i q_j}}{\hat{t}}=\dv{\hat{\sigma}_{\bar{q}_i \bar{q}_j\rightarrow \bar{q}_i \bar{q}_j}}{\hat{t}}=\dv{\hat{\sigma}_{q_i \bar q_j\rightarrow q_i \bar q_j}}{\hat{t}}=
	\frac{4}{9} \frac{\pi \alpha_{s}^{2} }{\hat s^{2}}\left ( \frac{\hat s^{2}+\hat u^{2}}{\hat t^{2}} \right ), 
	\\
	%-----------------------------------
	&\dv{\hat{\sigma}_{q_i g\rightarrow q_i g}}{\hat{t}}=\dv{\hat{\sigma}_{\bar q_i g\rightarrow \bar q_i g}}{\hat{t}}=
	\frac{\pi \alpha_{s}^{2} }{\hat{s}^{2}}\left [- \frac{4}{9}
	\left ( \frac{\hat{u}}{\hat{s}}+\frac{\hat{s}}{\hat{u}}
	\right ) +\frac{\hat{u}^{2}+\hat{s}^{2}}{\hat{t}^{2}} \right ],
	%-----------------------------------
	\end{align}
	%-----------------------------------
\end{subequations}
%-----------------------------------
where $\hat{s}$, $\hat{t}$ and  $\hat{u}$ are the partonic Mandelstam variables, and $q$ represents light quark.

Similarly, the inclusive photoproduction rate of the  $T_{4c}$ with large $p_{T}$ in $ep$ collisions can be written in the following form
%-----------------------------------
\bqa\label{t4c-cross section-ep}
%-----------------------------------
\mathrm{d} \sigma\left(e p \rightarrow T_{4c}\left(p_{\mathrm{T}}\right)+X\right) &=&\sum_{b,i}\int_0^1 \mathrm{d} x_{\gamma}  \int_0^{1} \mathrm{d} x_b \int_{0}^{1} \mathrm{d} z\; f_{\gamma/e}(x_{\gamma})f_{b/p}(x_b,\mu) \nn\\
&\times & \mathrm{d} {\hat \sigma}(\gamma+b \rightarrow i(p_T/z)+X, \mu)  D_{i \rightarrow T_{4c}}\left(z,\mu\right)+{\cal O}(1/p_T).
%------------------------------
\eqa
%-----------------------------------
The photon flux $f_{\gamma/e}$ is determined by equivalent
photon approximation (EPA)~\cite{Kniehl:1996we,Flore:2020jau}, which describes the possibility of finding a photon with a given momentum fraction in the electron: 
%-----------------------------------
\begin{align}
%-----------------------------------
& f_{\gamma / e}\left(x_{\gamma} \right)=\frac{\alpha}{2 \pi} {\left[\frac{1+\left(1-x_{\gamma}\right)^2}{x_{\gamma}} \ln \frac{Q_{\max }^2}{Q_{\min }^2\left(x_{\gamma}\right)}+2 m_e^2 x_{\gamma}\left(\frac{1}{Q_{\max }^2}-\frac{1}{Q_{\min }^2\left(x_{\gamma}\right)}\right)\right]},
%-----------------------------------
\end{align}
%-----------------------------------
where $Q_{\textrm{min}}^2 (x_{\gamma})=m_e^2x_{\gamma}^2/(1-x_{\gamma})$ and $m_e$ is the electron mass.
The value of $Q_{\max }^2$ varies across experiments, typically ranging around a few $\mathrm{GeV}^2$. 
The relevant partonic cross sections for the photoproduction processes at leading order are given by
%-----------------------------------
\begin{equation}
\begin{aligned}
\frac{d{\hat{\sigma}_{\gamma g\rightarrow q \bar q}}}{d\hat{t}}=\frac{2\pi \alpha \alpha_{s} e_{q}^{2}}{\hat s^{2}}
\left (\frac{\hat u}{\hat t}+\frac{\hat t}{\hat u}  \right ),
\end{aligned}
\end{equation}
%----------------------------------
and 
%-----------------------------------
\begin{equation}
\begin{aligned}
\frac{d{\hat{\sigma}_{\gamma q\rightarrow q g}}}{d\hat{t}}=\frac{d{\hat{\sigma}_{\gamma \bar q\rightarrow \bar q g}}}{d\hat{t}}= \frac{16}{3} \frac{\pi \alpha \alpha_{s} e_{q}^{2}}{\hat s^{2}}(-\frac{\hat t}{\hat s}-\frac{\hat s}{\hat t}),
\end{aligned}
\end{equation}
%----------------------------------
where $e_q$ denotes the charge of the light quark $q$. 

In 1982, Collins and Soper provided a gauge-invariant definition of the quark fragmentation function in $d=4-2\epsilon$ dimensions~\cite{Collins:1981uw}. For a light quark fragmenting into a hadron $H$, the fragmentation function is defined as
\begin{align} \label{t4c:Fragmentation:Function}
& D_{q \to H}(z,\mu) =
\frac{z^{d-3} }{ 2\pi \times 4 \times N_c }
\int_{-\infty}^{+\infty} \,dx^- \, e^{-i P^+ x^-/z}\nonumber
\\
& \times \text{tr}\left[ \hat{n}
\langle 0 | \Psi(0)
\Phi^\dagger(0,0,{\bf 0}_\perp) \sum_{X} |H(P)+X\rangle \langle H(P)+X|
\Phi (0,x^-,{\bf 0}_\perp) \bar{\Psi}(0,x^-,{\bf 0}_\perp) \vert 0 \rangle \right],
\end{align}
where $n^\mu =(0,1,{\bf 0}_\perp)$ is a null reference 4-vector and $\Psi$ denotes the initial quark field. Here, $z = P\cdot n / k\cdot n = P^{+} / k^{+}$ represents the light-cone momentum fraction, with $k$ being the momentum of the initial quark. The gauge link $\Phi (0,x^-,{\bf 0}_\perp)$ in the SU(3) adjoint ensures the gauge invariance of the fragmentation function.

% keep DGLAP?
The fragmentation function $D_{q\rightarrow T_{4c}}(z, \mu)$ obeys the celebrated
Dokshitzer-Gribov-Lipatov-Altarelli-Parisi (DGLAP) evolution equation~\cite{Altarelli:1977zs,Dokshitzer:1977sg,Gribov:1972ri,Lipatov:1974qm}:
%-----------------------------------
\beq\label{DGLAP:evolu:eq}
%-----------------------------------
\mu \frac{\partial}{\partial \mu} D_{q \rightarrow T_{4c}}(z, \mu)= {\frac{\alpha_s}{\pi}}\sum_{i\in\{q,g\}} \int_{z}^{1} \frac{\mathrm{d} y}{y} P_{iq}\left(\frac{z}{y}, \mu\right) D_{i \rightarrow T_{4c}}(y, \mu).
\eeq
%-----------------------------------
For concreteness, the quark-to-quark and quark-to-gluon splitting kernels are given at lowest order in $\alpha _{s}$ by
%-----------------------------------
\beq
%-----------------------------------
P_{qq}(z)=C_{F}\left [ \frac{1+z^{2}}{(1-z)_{+}}+\frac{3}{2}\delta (1-z) \right ],
%-----------------------------------
\eeq
%-----------------------------------
and
%-----------------------------------
{\beq
%-----------------------------------
% P_{q\rightarrow g}(z)=T_F \bqty{z^2+(1-z)^2}%C_{F}\frac{1+(1-z)^{2}}{z}.
P_{gq}(z)=C_{F}\frac{1+(1-z)^{2}}{z},
%-----------------------------------
\eeq
%-----------------------------------
where $C_F=4/3$.} % and $T_F={1}/{2}$. }

%-----------------------------
\section{NRQCD factorization for quark-to-$T_{4c}$ fragmentation\label{theoretical-framework}}
%-----------------------------
The fragmentation function of light hadrons is fundamentally a nonperturbative quantity, commonly accessed through experimental measurements or nonperturbative methods. {In contrast, the fragmentation functions for quarkonia or fully-heavy tetraquarks can be treated within the framework of QCD factorization.}
Prior to hadronization, the heavy quark and antiquark are first created at a short distance on the order of $\sim 1/m_c$. Asymptotic freedom allows us to decompose the fragmentation functions as the product of perturbatively calculable short-distance coefficients (SDCs) and nonperturbative long-distance matrix elements (LDMEs). {For a deeper understanding of the factorization in the fragmentation process, we refer the reader to some instructive literature~\cite{Suzuki:1977km,Cacciari:1993mq,Kniehl:2005mk}}.
In particular, the fragmentation functions for a light quark into a fully heavy hadron $H$ can be written as follows
%-----------------------------------
\beq
%-----------------------------------
D_{q \rightarrow H}(z)=\sum_{n} d_{n}(z)\left\langle {O}_{n}^{H}\right \rangle,
%-----------------------------------
\label{NRQCD:fac:quarkonium:frag}
%-----------------------------------
\eeq
%-----------------------------------
where $d_n(z)$ are SDCs and $\left\langle {O}_{n}^{H}\right \rangle$ signify various LDMEs. 

The fully-charmed $S$-wave tetraquarks, carrying the quantum number $0^{++}$ and $2^{++}$, are the main concern of this work. Notably, $C$-parity conservation prohibits the conversion of the $1^{+-}$ tetraquark from two gluons. 
In the diquark-antidiquark basis, the color-singlet tetraquark can be decomposed into either the ${\bar{\mathbf{3}}\otimes{\mathbf{3}}}$ or ${\mathbf{6}\otimes\bar{\mathbf{6}}}$ color configuration.
Fermi statistics dictates that the former case corresponds to the spin-1 diquark, while the latter corresponds to the spin-0 diquark. According to Eq.~\eqref{NRQCD:fac:quarkonium:frag}, the fragmentation function for a light quark into $T_{4c}$ at the lowest order in velocity can be expressed as
%-----------------------------------
\begin{align}
%-----------------------------------
\notag D_{q \rightarrow T_{4 c}}\left(z\right)=
&\frac{d_{3, 3}\left[q \rightarrow c c \bar{c} \bar{c}^{(J)}\right]}{m_{c}^{9}}
\left\langle {O}_{3,3}^{(J)}\right \rangle
+\frac{d_{6, 6}\left[q \rightarrow c c \bar{c} \bar{c}^{(J)}\right]}{m_{c}^{9}}
\left\langle {O}_{6,6}^{(J)}\right \rangle\\
&+\frac{d_{3, 6}\left[q \rightarrow c c \bar{c} \bar{c}^{(J)}\right]}{m_{c}^{9}}
2\mathrm{Re}[\left\langle {O}_{3,6}^{(J)}\right \rangle]+\cdots.
%-----------------------------------
\label{NRQCD:fac:t4c:fragmentation}
%-----------------------------------
\end{align}
%-----------------------------------
Here, the SDCs $d_n(z)$ are dimensionless. ${O}^{(J)}_{\rm color}$ ($J=0,2$) represent the NRQCD production operators with different color configuration, which are defined via
%-----------------------------------
\begin{subequations}
	%-----------------------------------
	\begin{align}
	%-----------------------------------
	&O^{J}_{3,3}=\mathcal{O}^{(J)}_{\bar{\mathbf{3}}\otimes{\mathbf{3}}} 
	\sum_{X}\left. | T^{J}_{4c}+X  \right. \rangle\left.\langle T^{J}_{4c}+X \right.|
	\mathcal{O}^{(J)\dagger }_{\bar{\mathbf{3}}\otimes{\mathbf{3}}},
	%-----------------------------------
	\\
	%-----------------------------------
	&O^{J}_{6,6}=\mathcal{O}^{(J)}_{{\mathbf{6}}\otimes \bar{\mathbf{6}}} 
	\sum_{X}\left. | T^{J}_{4c}+X  \right. \rangle\left.\langle T^{J}_{4c}+X \right.|
	\mathcal{O}^{(J)\dagger }_{{\mathbf{6}}\otimes \bar{\mathbf{6}}},
	%-----------------------------------
	\\
	%-----------------------------------
	&O^{J}_{3,6}=\mathcal{O}^{(J)}_{\bar{\mathbf{3}}\otimes{\mathbf{3}}} 
	\sum_{X}\left. | T^{J}_{4c}+X  \right. \rangle\left.\langle T^{J}_{4c}+X \right.|
	\mathcal{O}^{(J)\dagger }_{{\mathbf{6}}\otimes \bar{\mathbf{6}}},
	%-----------------------------------
	\end{align}
	%-----------------------------------
\end{subequations}
%-----------------------------------
with the composite color-singlet four-quark operators $\mathcal {O}^{(J)}_{\rm color}$. These composite operators are expressed by~\cite{Huang:2021vtb,Feng:2020riv} 
%-----------------------------------
\begin{subequations}
	%-----------------------------------
	\begin{align}
	%-----------------------------------
	&\mathcal{O}^{(0)}_{\bar{\mathbf{3}}\otimes{\mathbf{3}}}=-\frac{1}{\sqrt{3}}[\psi_a^T(i\sigma^2)\sigma^i\psi_b] [\chi_c^{\dagger}\sigma^i (i\sigma^2)\chi_d^*]\;
	\mathcal{C}^{ab;cd}_{\bar{\mathbf{3}}\otimes{\mathbf{3}}},
	%-----------------------------------
	%\\
	%-----------------------------------
	%& \mathcal{O}^{(1)}_{\bar{\mathbf{3}}\otimes{\mathbf{3}}}= -{\frac{i}{\sqrt{2}}} \left[\psi_a^T (i \sigma^2)\sigma^j\psi_b\right]\left[\chi_c^\dagger\sigma^k (i \sigma^2)\chi_d^*\right]\,\epsilon^{ijk}\;{\mathcal C}^{ab;cd}_{\bar{\mathbf{3}}\otimes{\mathbf{3}}},
	%-----------------------------------
	\\
	%-----------------------------------
	&\mathcal{O}^{(2)}_{\bar{\mathbf{3}}\otimes{\mathbf{3}}}=\frac{1}{2}[\psi_a^T(i\sigma^2)\sigma^m\psi_b] [\chi_c^{\dagger}\sigma^n(i\sigma^2)\chi_d^*]\;\Gamma^{ij;mn}
	\;\mathcal{C}^{ab;cd}_{\bar{\mathbf{3}}\otimes{\mathbf{3}}},
	%-----------------------------------
	\\
	%-----------------------------------
	&\mathcal{O}^{(0)}_{\mathbf{6}\otimes\bar{\mathbf{6}}}=
	[\psi_a^T(i\sigma^2)\psi_b] [\chi_c^{\dagger}(i\sigma^2)\chi_d^*]\;
	\mathcal{C}^{ab;cd}_{\mathbf{6}\otimes\bar{\mathbf{6}}},
	%-----------------------------------
	\end{align}
	%-----------------------------------
	\label{NRQCD:composite:operators}
	%-----------------------------------
\end{subequations}
%-----------------------------------
where $\sigma^i$ is Pauli matrix, $\psi$ and $\chi^\dagger$ are Pauli spinor fields.
The rank-4 symmetric traceless tensor is given by $\Gamma^{ij;mn}\equiv \delta^{i m} \delta^{j n}+\delta^{i n} \delta^{j m}-\frac{2}{3} \delta^{i j} \delta^{mn}$,
and the color projection tensors read
%-----------------------------------
\begin{subequations}
	%-----------------------------------
	\bqa
	%-----------------------------------
	&& \mathcal{C}^{ab;cd}_{\bar{\mathbf{3}}\otimes{\mathbf{3}}}\equiv \frac{1}{2\sqrt{3}}(\delta^{ac}\delta^{bd}-\delta^{ad}\delta^{bc}),
	%-----------------------------------
	\\
	%-----------------------------------
	&& \mathcal{C}^{ab;cd}_{{\mathbf{6}}\otimes \bar {\mathbf{6}}}
	\equiv \frac{1}{2\sqrt{6}}(\delta^{ac}\delta^{bd}+\delta^{ad}\delta^{bc}).
	%-----------------------------------
	\eqa
	%-----------------------------------
	\label{color:tensor}
	%-----------------------------------
\end{subequations}
%-----------------------------
%-----------------------------

In the spirit of factorization, the SDCs in Eq.~\eqref{NRQCD:fac:t4c:fragmentation} are insensitive to the long-distance dynamics within the tetraquark. Consequently, they can be determined by replacing the physical tetraquark state with a fictitious state consisting of four free charm quarks. By computing both sides of Eq.\eqref{NRQCD:fac:t4c:fragmentation} perturbatively, the SDCs can be extracted. This procedure follows standard perturbative matching techniques as outlined in~\cite{Feng:2020riv,Feng:2020qee,Bai:2024ezn}. In these calculations, the following covariant projectors were employed to project out amplitudes with desired spin/color quantum numbers
%-----------------------------------
\beq
%-----------------------------------
\bar u^a_i\bar u^b_j v^c_k v^d_l\to (\textsf{C}\Pi_\mu)^{ij}(\Pi_\nu \textsf{C} )^{lk}\mathcal{C}^{ab;cd}_{\rm color} J^{\mu\nu}_{0,2},
%-----------------------------------
\eeq
%-----------------------------------
where $\textsf{C}=i\gamma^2\gamma^0$ is the charge conjugation matrix, $a$, $b$, $c$ and $d$ denote the color indices, and $i$, $j$, $k$ and $l$ denote the spinor indices. $\Pi_\mu$ is the spin-triplet projector for two fermion, as defined in~\cite{Petrelli:1997ge}. The covariant projectors $J^{\mu\nu}_{0,2}$ are defined as
%-----------------------------------
\begin{subequations}
	%-----------------------------------
	\begin{align}
	%-----------------------------------
	&J^{\mu \nu }_{0}=\frac{1}{\sqrt{3}}\eta ^{\mu \nu},
	%-----------------------------------
	\\
	%-----------------------------------
	&J^{\mu \nu }_{2}(\epsilon )=\epsilon_
	{\alpha \beta}\left \{ \frac{1}{2}\left [ \eta^{\mu \alpha} \eta^{\nu \beta} +\eta^{\mu \beta} \eta^{\nu \alpha} \right ]
	-\frac{1}{3}\eta^{\mu \nu}\eta^{\alpha \beta}  \right \},
	%-----------------------------------
	\end{align}
	%-----------------------------------
	\label{C-G:coupling}
	%-----------------------------------
\end{subequations}
%-----------------------------------
where $p$ denotes the momentum of $T_{4c}$ and $\eta^{\mu \nu}\equiv -g^{\mu \nu}+p^{\mu}p^{\nu}/p^{2}$. 

%-----------------------------------
\begin{figure}[hbtp]
	\centering
	\includegraphics[width=0.5\textwidth]{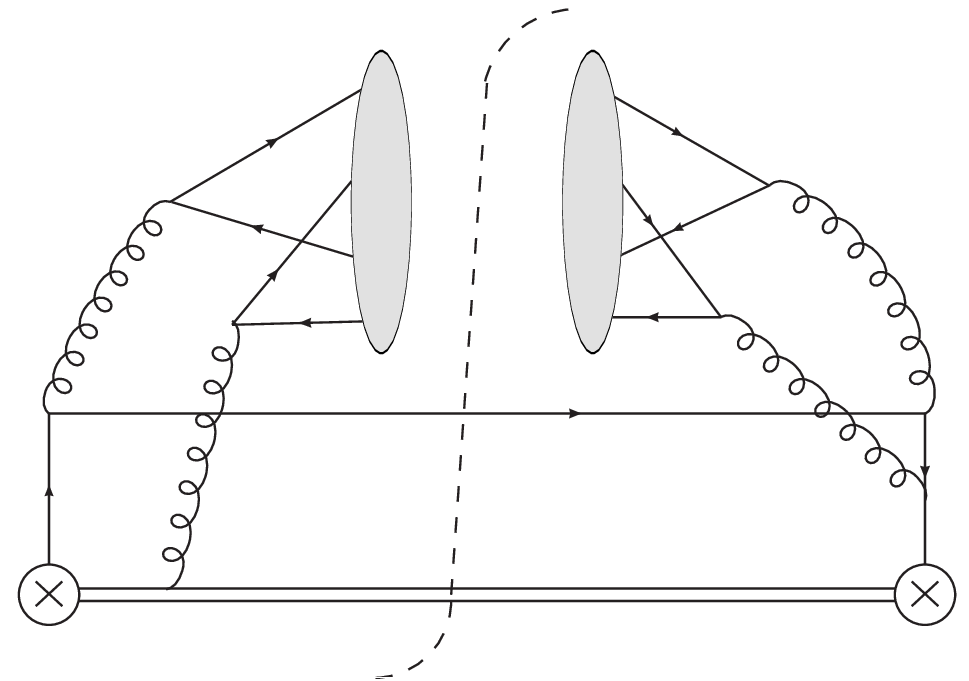}
	\caption{Typical Feynman diagram for the fragmentation function of a light quark into $T_{4c}$, drawn with {\tt JaxoDraw}~\cite{Binosi:2008ig}. Horizontal double line represent the eikonal gauge link and the grey blob denotes the tetraquark state.
}
	\label{diagrams}
\end{figure}
%-----------------------------------

%-----------------------------------
\section{Determination of short-distance coefficients\label{determination-sdcs} }
%-----------------------------------
We use the the package {\tt FeynArts}~\cite{Hahn:2000kx} to generate the Feynman diagrams and the corresponding amplitudes for the partonic fragmentation function $D_{q\to cc\bar{c}\bar{c}}$. 
At the lowest order in $\alpha_{s}$, there are 32 diagrams for each side of the cut. A typical Feynman diagram is shown in Fig.~\ref{diagrams}. 
The Dirac trace algebra and Lorentz contraction are 
handled using the packages {\tt FeynCalc/FormLink}~\cite{Mertig:1990an,Feng:2012tk}.

The SDCs are determined using the technique outlined in the previous section.
For the fragmentation of a light quark into a $0^{++}$ tetraquark, we obtain the following results for SDCs in~\eqref{NRQCD:fac:t4c:fragmentation}
%-----------------------------------
%-----------------------------------

\begin{subequations}\label{eq:sdc-0}
	\begin{align}
	&d_{6,6}\left(q \rightarrow 0^{++}\right)=\frac{\pi^{2}\alpha_{s}^{4}}{1728 m_{c}^{4} z}\,\mathcal{G},\\
 %    \begin{aligned}[t]
	% &  \frac{\pi^{2}\alpha_{s}^{4}}{1728 m_{c}^{4} z}\bigg\{m_{q}^{2}\left(2m_{c}^{2}(z-4)+m_{q}^{2}z\right) \log\left[\frac{8m_{c}^{2}(z-2)(z-1)+m_{q}^{2} z^2}{m_{q}^{2} z^2-16m_{c}^{2} (z-1)}\right]\\ 
 %    &+\frac{8m_{c}^{2}(z-1)}{(16m_{c}^{2}(z-1)-m_{q}^{2}z^2)(8m_{c}^{2}(z-2)(z-1)+m_{q}^{2}z^2)}\\
	% &\times \left[32m_{c}^{6}(z^2-5z+4)^2-2m_{c}^{4} m_{q}^{2}(z-1)z(z^3+32z-64)\right.\\
	% &\left.+2m_{c}^{2} m_{q}^{4}(z-4)z^2(3z-2)+m_{q}^{6}z^4\right]\bigg\},
	% \end{aligned}\\
	&d_{3, 3}\left(q \rightarrow  0^{++}\right)= \frac{\pi^{2}\alpha_{s}^{4}}{2592m_c^4 z}\,\mathcal{G},\\
 %    \begin{aligned}[t]
	% & \frac{\pi^{2}\alpha_{s}^{4}}{2592m^4 z}\bigg\{m_{q}^{2}\left(2m_{c}^{2}(z-4)+m_{q}^{2}z\right) 
	% \log\left[\frac{8m_{c}^{2}(z-2)(z-1)+m_{q}^{2} z^2}{m_{q}^{2} z^2-16m_{c}^{2} (z-1)}\right]\\
	% &+\frac{8m_{c}^{2}(z-1)}{(16m_{c}^{2}(z-1)-m_{q}^{2}z^2)(8m_{c}^{2}(z-2)(z-1)+m_{q}^{2}z^2)}\\
	% &\times \left[32m_{c}^{6}(z^2-5z+4)^2-2m_{c}^{4} m_{q}^{2}(z-1)z(z^3+32z-64)\right.\\
	% &\left.+2m_{c}^{2} m_{q}^{4}(z-4)z^2(3z-2)+m_{q}^{6}z^4\right]\bigg\},\\
	% \end{aligned}\\
	&d_{3,6}\left(q \rightarrow 0^{++}\right)= \frac{\pi^{2}\alpha_{s}^{4}}{864 \sqrt{6} m_c^4  z}\,\mathcal{G},
 %    \begin{aligned}[t]
	% & \frac{\pi^{2}\alpha_{s}^{4}}{864 \sqrt{6} m^4  z}\bigg\{m_{q}^{2}\left(2m_{c}^{2}(z-4)+m_{q}^{2}z\right) 
	% \log\left[\frac{8m_{c}^{2}(z-2)(z-1)+m_{q}^{2} z^2}{m_{q}^{2} z^2-16m_{c}^{2} (z-1)}\right]\\
	% &+\frac{8m_{c}^{2}(z-1)}{(16m_{c}^{2}(z-1)-m_{q}^{2}z^2)(8m_{c}^{2}(z-2)(z-1)+m_{q}^{2}z^2)}\\
	% &\times \left[32m_{c}^{6}(z^2-5z+4)^2-2m_{c}^{4} m_{q}^{2}(z-1)z(z^3+32z-64)\right.\\
	% &\left.+2m_{c}^{2} m_{q}^{4}(z-4)z^2(3z-2)+m_{q}^{6}z^4\right]\bigg\},\\
	% \end{aligned}
	\end{align}
\end{subequations}
%-----------------------------------
 where $m_{q}$ denotes the mass of light quark, and 
\begin{align}
    \mathcal{G}=\begin{aligned}[t]
    &m_{q}^{2}\left(2m_{c}^{2}(z-4)+m_{q}^{2}z\right) \log\left[\frac{8m_{c}^{2}(z-2)(z-1)+m_{q}^{2} z^2}{m_{q}^{2} z^2-16m_{c}^{2} (z-1)}\right]\\ 
    &+\frac{8m_{c}^{2}(z-1)}{(16m_{c}^{2}(z-1)-m_{q}^{2}z^2)(8m_{c}^{2}(z-2)(z-1)+m_{q}^{2}z^2)}\left[32m_{c}^{6}(z^2-5z+4)^2\right.\\
	&-2m_{c}^{4} m_{q}^{2}(z-1)z(z^3+32z-64)\left.+2m_{c}^{2} m_{q}^{4}(z-4)z^2(3z-2)+m_{q}^{6}z^4\right].
    \end{aligned}
\end{align}

The SDC for the fragmentation of a light quark into a $2^{++}$ tetraquark is given as follows:
%-----------------------------------
\begin{equation}\label{eq:sdc-2}
\begin{aligned}
& d_{3,3}\left(q \rightarrow 2^{++}\right)= \frac{\pi^{2}\alpha_{s}^{4}}{3240 m_{c}^{4} z^2}\bigg\{-\left(12m_{c}^{4}((z-2)^2z-4)-m_{c}^{2} m_{q}^{2}(z-1)z(3z-16)+2m_{q}^{4}z^2\right) \\
% &\times\left(\log\left[m_{q}^{2} z^2-16 m_{c}^{2}(z-1)\right]-\log\left[8m_{c}^{2}(z-2)(z-1)+m_{q}^{2}z^2\right]\right)\\
&\times \log\left[\frac{m_{q}^{2} z^2-16 m_{c}^{2}(z-1)}{8m_{c}^{2}(z-2)(z-1)+m_{q}^{2}z^2}\right]
+\frac{8m_{c}^{2}(z-1)z}{(16m_{c}^{2}(z-1)-m_{q}^{2}z^2)(8m_{c}^{2}(z-2)(z-1)+m_{q}^{2}z^2)}\\
&\times \left[16m_{c}^{6}(z-1)(z((z-33)z+72)-16)+m_{c}^{4} m_{q}^{2}z(z(z((97-13z)z-440)\right.\\
&\left.+576)-256)-m_{c}^{2}m_{q}^{4}(z-1)z^2(3(z-8)z+32)+2m_{q}^{6}z^4\right]\bigg\}.
\end{aligned}
\end{equation}
%-----------------------------------

It is important to note that we have temporarily retained the mass of the light quark $m_q$, 
in Eqs (\ref{eq:sdc-0}) and (\ref{eq:sdc-2}). In fact, we can derive the fragmentation function for a bottom quark to $T_{4c}$ by substituting $m_q$ with $m_b$. Similarly, we can derive the fragmentation function for a charm quark to $T_{4b}$ by making the substitutions $m_c \to m_b$ and $m_q \to m_c$ in Eqs. (\ref{eq:sdc-0}) and (\ref{eq:sdc-2}).

{It is worth emphasizing that in Eqs. (\ref{eq:sdc-0}) and (\ref{eq:sdc-2}), the typical energy scale for the strong coupling constant should be around $\mu=4m_c$}.

For the physical scenario involving the fragmentation of a light quark, we must set $m_q=0$ in (\ref{eq:sdc-0}) and (\ref{eq:sdc-2}), which yields
\begin{subequations}
	%-------------------------
	\begin{align}
	%-------------------------
	& d_{6,6}\left(q \rightarrow 0^{++}\right) =\frac{\pi^{2}\alpha_{s}^{4} (z-4)^{2}(z-1)}{864z (z-2)},
	%-------------------------
	\\
	& d_{3,3}\left(q \rightarrow 0^{++}\right) =\frac{\pi^{2}\alpha_{s}^{4} (z-4)^{2}(z-1)}{1296z (z-2)},
	%-------------------------
	\\
	%-------------------------
	& d_{3,6}\left(q \rightarrow 0^{++}\right) =\frac{\pi^{2}\alpha_{s}^{4} (z-4)^{2}(z-1)}{432\sqrt{6} z (z-2)},
	%-------------------------
	%\\
	%-------------------------
	%& d_{3,3}\left(q \rightarrow 1^{+-}\right) =0,
	%-------------------------
	\\
	%-------------------------
	& d_{3,3}\left(q \rightarrow 2^{++}\right) =
 \begin{aligned}[t]
     &\frac{\pi^{2}\alpha_{s}^{4}}{3240 z^2 (z-2)}\bigg[z(z^3-33z^2+72z-16)\\
	&-12(z^4-6z^3+12z^2-12z+8) \log(\frac{2}{2-z})\bigg].
 \end{aligned}
	%-------------------------
	\end{align}
	%-------------------------
	\label{limit-behavior}
	%-------------------------
\end{subequations}
%-------------------------

%--------------------------------------------------------------------
\section{Phenomenology and discussion\label{phen-discu}}
%-----------------------------------
To make phenomenological predictions, it is crucial to first determine the specific values of the LDMEs that appear in the factorization formula \eqref{NRQCD:fac:t4c:fragmentation}. Utilizing the vacuum saturation approximation, the NRQCD LDMEs can be expressed in terms of the wave function of the $T_{4c}$ at the origin~\cite{Feng:2023agq}
%-----------------------------------
\begin{subequations}
	%-------------------------
	\begin{align}
	%-------------------------
	& \left\langle {O}_{C_1,C_2}^{(0)}\right \rangle \approx 16\,{\psi_{C_1}(\mathbf{0})\psi_{C_2}^*(\mathbf{0})},
	%-------------------------
	% \\
	% & \left\langle {O}_{C_1,C_2}^{(1)}\right \rangle \approx 48\,{\psi_{C_1}(\mathbf{0})\psi_{C_2}^*(\mathbf{0})},
	% %-------------------------
	\\
	%-------------------------
	& \left\langle {O}_{C_1,C_2}^{(2)}\right \rangle \approx 80\,{\psi_{C_1}(\mathbf{0})\psi_{C_2}^*(\mathbf{0})},
	%-------------------------
	\end{align}
	%-------------------------
	\label{LDMEs}
	%-------------------------
\end{subequations}
%-------------------------
where $\psi(\mathbf{0})$ denotes the four-body Schr\"odinger wave function at the origin. The color structure labels, $C_1$ and $C_2$, can assume values of $3$ or $6$, which correspond to the diquark-antidiquark configurations $\bar{\mathbf{3}} \otimes \mathbf{3}$ and $\mathbf{6} \otimes \bar{\mathbf{6}}$, respectively. 
In phenomenology, utilizing various potential models, extensive research has been conducted on the four-body Schr\"odinger wave functions for the fully-charmed tetraquarks.\footnote{It is suggested that the experimentally observed $X(6900)$ is more likely to be the $2S$ radial excited state ~\cite{Lu:2020cns,Zhao:2020nwy,liu:2020eha,Yu:2022lak,Wang:2021kfv,Liu:2021rtn,Belov:2024qyi}. The values of the LDMEs from five distinct  models~\cite{Lu:2020cns,Zhao:2020nwy,liu:2020eha,Yu:2022lak,Wang:2019rdo}, including radially excited states up to the $3S$ level, are enumerated in~\cite{Bai:2024ezn}.}  
In this work, we adopt the wavefunction at the origin from \cite{Lu:2020cns} to estimate the LDMEs, which are given in Table.~\ref{tab:LDME}. 

%-------------------------
\begingroup
%-------------------------
\setlength{\tabcolsep}{14pt} % Default value: 6pt
%-------------------------
\renewcommand{\arraystretch}{2}
%-------------------------
\begin{table}[H]
	%-------------------------
	\centering
	%-------------------------
	\begin{tabular}{cccc}
		\hline
		\hline
		&\multicolumn{1}{c}{\textbf{LDME}}$[\mathrm{GeV}^9]$ & \textbf{1S} & \textbf{2S} \\
		\hline
		\multirow{3}{*}{$0^{++}$} & $\expval{{O}_{6,6}^{(0)}}$
		& $0.0128$ & $0.0347$ \\
		\cline{2-4}
		&$\expval{{O}_{3,6}^{(0)}}$ & $0.0211$ & $0.0538$ \\
		\cline{2-4}
		&$\expval{{O}_{3,3}^{(0)}}$ & $0.0347$ & $0.0832$ \\
		\hline
		$2^{++}$&$\expval{{O}_{3,3}^{(2)}}$ & $0.072$ & $0.1775$ \\
		\hline
		\hline
	\end{tabular}
	\caption{Numerical values of the LDMEs estimated from \cite{Lu:2020cns}.}
	%-------------------------
	\label{tab:LDME}
	%-------------------------
\end{table}
%-------------------------
\endgroup
%-------------------------

\begin{figure}[H]
	\centering
	\includegraphics[width=0.49\linewidth]{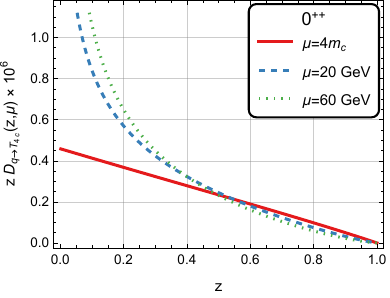}
	\includegraphics[width=0.49\linewidth]{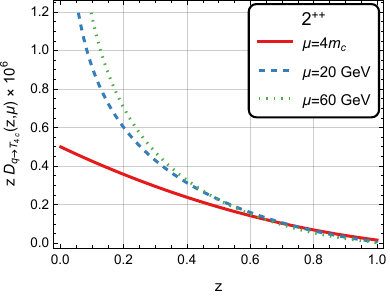}
	\caption{Fragmentation functions $D_{q\to T_{4c}}(z,\mu)$ at initial scale $\mu=4m_c$ (solid red) and after evolution to $\mu=20\ \mathrm{GeV}$ (dashed blue) and $\mu=60\ \mathrm{GeV}$ (dotted green) for $0^{++}$ (left) and $2^{++}$ tetraquarks (right). The LDMEs are taken from the $1S$ tetraquarks in Table~\ref{tab:LDME}. }
	\label{fig:fragmentation}
\end{figure}

{
Using the LDMEs of $1S$ tetraquarks from Table~\ref{tab:LDME}, we calculate the fragmentation functions $D_{q\to T_{4c}}(z,\mu)$ at the initial scale $\mu = 4m_c$, with $m_c = 1.5\ \mathrm{GeV}$. 
We then compute the evolution of the fragmentation functions by numerically solving the DGLAP equations using the Runge-Kutta method, incorporating contributions from all parton channels. The results are presented in Fig.~\ref{fig:fragmentation}, showing the fragmentation functions at the initial scale as well as at $\mu = 20\ \mathrm{GeV}$ and $\mu = 60\ \mathrm{GeV}$. 
 
We notice that the evolution of the light quark fragmentation functions has a significant impact in the small- and mid-$z$ regions, primarily due to mixing with the gluon fragmentation functions. For both the \(0^{++}\) and \(2^{++}\)  
states, the fragmentation functions increase notably at small \(z\) as \(\mu\)  
increases, while the large-\(z\) region experiences a slight decrease. }

%-------------------------------------------------------
%------------------------------
\subsection{$p_T$ distribution for the inclusive production of $T_{4c}$ at the LHC}
%------------------------------
To determine the $p_T$ distributions of $T_{4c}$ at the \texttt{LHC}, we apply the cross section formula given in Eq.~\eqref{t4c-cross section}. We assume a center-of-mass (CM) energy of $\sqrt{s} = 13$ TeV and set the charm quark mass to be $m_c = 1.5$ $\rm GeV$. The factorization scale is chosen to be $\mu = M_T$, where $M_T = \sqrt{p_T^2 + M^2_{T_{4c}}}$ denotes the transverse mass of $T_{4c}$. To account for scale uncertainties, we vary $\mu$ within the range $M_T/2 \leq \mu \leq 2 M_T$. A rapidity cut of $-5 \leq y \leq 5$ is also applied. In addition, we utilize the \texttt{CT14llo PDF} sets~\cite{Dulat:2015mca}.
%{\color{red} Evidence is provided that, when the production rate of $T_{4c}$ is detected with $p_{T}$ exceeding 20GeV, fragmentation mechansim may begin to predominate as indicated in~\cite{}.}

\begin{figure}[H]
	\centering
	\includegraphics[width=0.49\linewidth]{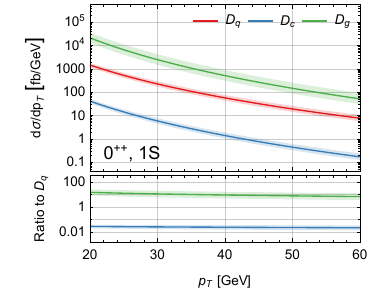}
	\includegraphics[width=0.49\linewidth]{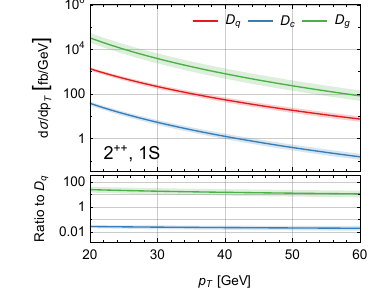}\\
	\includegraphics[width=0.49\linewidth]{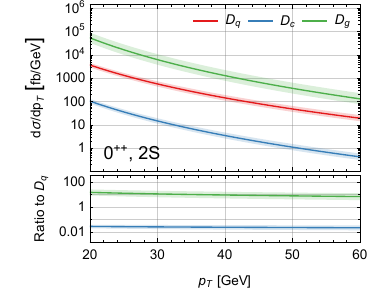}
	\includegraphics[width=0.49\linewidth]{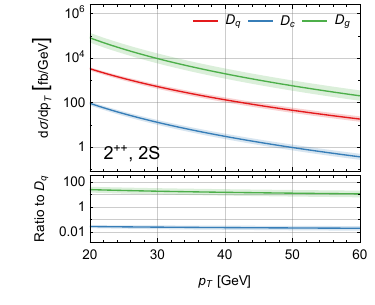}
	\caption{The $p_T$ distributions of $T_{4c}$ production at the \texttt{LHC} from light quark fragmentation (denoted as $D_q$) compared to gluon ($D_g$) and charm quark ($D_c$) fragmentations. The lower insets show the ratios of the gluon/charm quark fragmentation to the light quark fragmentation.}
	\label{fig:LHC_ufrag_comp_pT_dist}
\end{figure}

Fig.~\ref{fig:LHC_ufrag_comp_pT_dist} compares the $p_{T}$ distributions of $T_{4c}$ production from light quark fragmentation to those from gluon~\cite{Feng:2020riv} and charm quark fragmentation~\cite{Bai:2024ezn}, focusing on $p_T > 20$ GeV. We note that all $p_{T}$-distributions follow a common trend: they fall down as $p_{T}$ increases.
Overall, the $p_{T}$ distributions of $T_{4c}$ production from light quark fragmentation contributions are approximately 1-2 orders of magnitude smaller than those from gluon fragmentation, while being 1-2 orders of magnitude larger than those from charm quark fragmentation. Additionally, the differential cross sections for the distinct $T_{4c}$ states exhibit similar $p_{T}$-dependent trends across all three fragmentation channels.

%------------------------------
\subsection{$p_T$ distribution for inclusive production of $T_{4c}$ at the EIC}
%------------------------------
To study light quark fragmentation contributions at electron-ion colliders, we focus on the \texttt{EIC} with a CM energy of $140\ \mathrm{GeV}$. This setup, among the various energy configurations of the EIC, as well as compared to \texttt{HERA} and \texttt{EicC}, is the most promising for observing the $T_{4c}$~\cite{Feng:2023ghc}. We follow the same approach as in the previous section, applying cuts on the elasticity parameter, $0.05 < z < 0.9$, and on the photon virtuality, $Q_{\tt max}^2 = 1\ \mathrm{GeV}^2$, instead of the tetraquark rapidity. The electron mass is taken as $0.51 \times 10^{-3}\ \mathrm{GeV}$, and the fine structure constant is $\alpha = 1/137$.

\begin{figure}[H]
	\centering
	\includegraphics[width=0.49\linewidth]{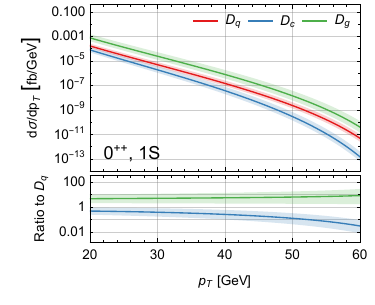}
	\includegraphics[width=0.49\linewidth]{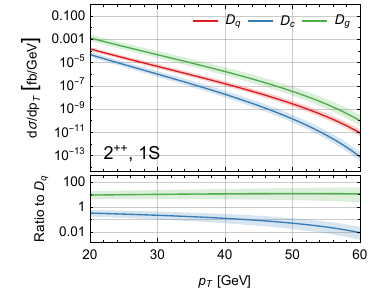}\\
	\includegraphics[width=0.49\linewidth]{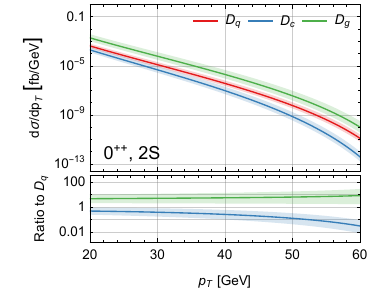}
	\includegraphics[width=0.49\linewidth]{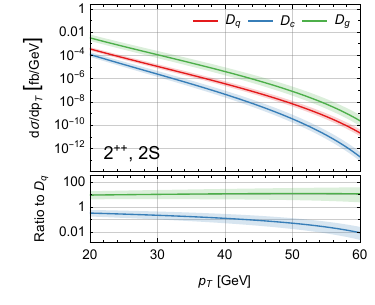}
	\caption{The $p_T$ distributions of $T_{4c}$ production at the \texttt{EIC} from light quark fragmentation (denoted as $D_q$) compared to gluon ($D_g$) and charm quark ($D_c$) fragmentation. The lower insets show the ratio of the gluon/charm quark fragmentation to the light quark fragmentation.}
	\label{fig:EIC_ufrag_comp_pT_dist}	
\end{figure}

Fig.~\ref{fig:EIC_ufrag_comp_pT_dist} presents a comparison between the contributions of light quark fragmentation and those of gluon and charm quark fragmentation to the $p_T$ distribution at the \texttt{EIC}~\cite{Bai:2024ezn}. We find that, while light quark fragmentation contributions are comparable to charm quark fragmentation at low $p_T$, they decrease much more slowly as $p_T$ increases. Both quark fragmentation contributions are at least an order of magnitude smaller than the gluon fragmentation, which also shows the slowest decrease with $p_T$. The $p_{T}$ scaling behavior is similar across all three fragmentation channels.

\subsection{Integrated cross sections}
To further quantify the contributions of light quark fragmentation, we calculate the integrated cross sections for $T_{4c}$ production at the \texttt{LHC} and \texttt{EIC}. The results are summarized in Table~\ref{tab:cross-section}. Comparing these with the integrated cross sections from gluon~\cite{Feng:2020riv} and charm quark fragmentation~\cite{Bai:2024ezn}, we confirm the conclusions from previous sections. At the \texttt{LHC}, the cross sections for $T_{4c}$ production from light quark fragmentation is significantly smaller than that from gluon fragmentation but remains more substantial than that from charm quark fragmentation.
This significant difference can be explained by the following factors. 
Firstly, across most $p_{T}$ regimes, the partonic cross sections specified in (\ref{partonic:cross:section}) are considerably smaller than those for $gg \to gg$; on the other hand, contributions from light quark fragmentation involves a greater number of partonic channels compared to charm quark fragmentation. 
Secondly, the magnitude of the fragmentation function for $q\to T_{4c}$ and $c\to T_{4c}$ is notably smaller than that for $g\to T_{4c}$.
At the \texttt{EIC}, the contribution of light quark fragmentation to the integrated cross sections is comparable to that of charm quark fragmentation, though gluon fragmentation remains the dominant contribution. 
It is observed that the cross sections for $T_{4c}$ production at \texttt{EIC} are significantly smaller than those at the \texttt{LHC}.

\begingroup
%-------------------------
\setlength{\tabcolsep}{8pt} % Default value: 6pt
%-------------------------
\renewcommand{\arraystretch}{2}
\begin{table}[h!]
	\centering
	\begin{tabular}{cccccc}
		\hline
		\hline
		&\multirow{2}*{$\bm{nS}$} & \multirow{2}*{$\bm{J^{PC}}$} & \multicolumn{3}{c}{\textbf{Integrated cross section [fb]}} \\
		\cline{4-6}
		& & & $D_q$ & $D_g$ & $D_c$ \\
		\hline
		\multirow{4}*{\texttt{LHC}} & \multirow{2}*{$1S$} & $0^{++}$ & $8.1 \times 10^3$ & $1.1 \times 10^5$ & $2.2 \times 10^2$\\
		&                                              & $2^{++}$ & $7.5 \times 10^3$ & $1.7 \times 10^5$ & $2.0 \times 10^2$\\
		\cline{2-6}
		& \multirow{2}*{$2S$} & $0^{++}$ & $2.0 \times 10^4$ & $2.9 \times 10^5$ & $5.5 \times 10^2$\\
		&                                              & $2^{++}$ & $1.8 \times 10^4$ & $4.3 \times 10^5$ & $4.8 \times 10^2$\\
		\hline
		\multirow{4}{*}{\texttt{EIC}}& \multirow{2}*{$1S$} & $0^{++}$ & $5.0 \times 10^{-4}$ & $2.7 \times 10^{-3}$ & $2.4 \times 10^{-4}$ \\
		&                                                 & $2^{++}$ & $4.4 \times 10^{-4}$ & $4.7 \times 10^{-3}$ & $1.4 \times 10^{-4}$ \\
		\cline{2-6}
		& \multirow{2}{*}{$2S$} & $0^{++}$ & $1.3 \times 10^{-3}$ & $6.7 \times 10^{-3}$ & $6.0 \times 10^{-4}$\\
		&                                              & $2^{++}$ & $1.1 \times 10^{-3}$ & $1.2 \times 10^{-2}$ & $3.4 \times 10^{-4}$\\
		\hline
		\hline
	\end{tabular}
	\caption{Integrated cross section with $p_T \geq 20\ \mathrm{GeV}$ for $0^{++}$ and $2^{++}$ tetraquarks at the \texttt{LHC} and \texttt{EIC}. The contributions from light quark fragmentation (denoted as $D_q$) are compared to those from gluon ($D_g$)~\cite{Feng:2020riv,Bai:2024ezn} and charm quark ($D_c$)~\cite{Bai:2024ezn} fragmentation.}
	\label{tab:cross-section}
\end{table}

\endgroup

%---------------------------
\section{Summary\label{summary}}
%---------------------------

In this paper, within the NRQCD factorization framework, we calculate the fragmentation function for a light quark into an $S$-wave fully-charmed tetraquark. We compute the SDCs for the fragmentation of a light quark into a $0^{++}$ and $2^{++}$ tetraquark at the lowest order in $\alpha_{s}$. 
The nonperturbative LDMEs are evaluated from phenomenological potential model, with the aid of vacuum saturation approximation.
We present the $p_T$ distributions for $T_{4c}$ production from light quark fragmentation, and compare them to those from gluon and charm quark fragmentation, while also predicting the $T_{4c}$ production cross sections in the high-$p_T$ region at the \texttt{LHC} and \texttt{EIC}. 
The cross section for $T_{4c}$ production from light quark fragmentation is considerablely smaller than that from gluon fragmentation, yet remains larger than that from charm quark fragmentation.
% NOTE: say something about the production prosect or not? it's only one channel
% Our study suggests that a considerable potential for $T_{4c}$ event production at the  \texttt{LHC}, yet the prospect of detecting these $T_{4c}$ at the \texttt{EIC} is gloomy.

\section*{Acknowledgments}
%---------------------
The work of X.-W. B., and W.-L. S. is supported by the
National Natural Science Foundation of China under Grants 
No. 12375079, and the Natural Science
Foundation of ChongQing under Grant No. CSTB2023
NSCQ-MSX0132. The work of Y.-S.~H. is supported by the DOE grants DE-FG02-91ER40684 and DE-AC02-06CH11357.
%---------------------------------

\end{document}